# Acene:spacer blends show exothermic singlet fission to proceed coherently, while endothermic fission proceeds incoherently

Clemens Zeiser[1], Chad Cruz[2], David R. Reichman[3], Michael Seitz[4], Jan Hagenlocher[1], Eric L. Chronister[5], Christopher Bardeen[2], Roel Tempelaar[3,6], and Katharina Broch[1]

[1]Institute of Applied Physics, University of Tübingen, Auf der Morgenstelle 10, 72076 Tübingen, Germany
[2]Department of Chemistry, University of California at Riverside, 501 Big Springs Road, Riverside, CA 92521, USA
[3]Department of Chemistry, Columbia University, 3000 Broadway, New York, NY 10027, USA
[4]Institute of Inorganic Chemistry, University of Tübingen, Auf der Morgenstelle 18, 72076 Tübingen, Germany
[5]University of Nevada, Las Vegas, 4505 S. Maryland Pkwy., Las Vegas, NV 89154, USA
[6]Department of Chemistry, Northwestern University, 2145 Sheridan Rd, Evanston, IL 60208, USA

## Abstract

The fission of singlet excitons into triplet pairs in organic materials holds great technological promise, but the rational application of this phenomenon is hampered by a lack of understanding of its complex photophysics. Here, we use the controlled introduction of vacancies by means of spacer molecules in tetracene and pentacene thin films as a tuning parameter complementing experimental observables to identify the operating principles of different singlet fission pathways. Time-resolved spectroscopic measurements in combination with microscopic modelling enables us to demonstrate distinct scenarios, resulting from different singlet-to-triplet energy alignments. For exothermic fission, as found for pentacene, coherent mixing between the photoexcited singlet and triplet pair states is promoted by vibronic resonances, which drives the fission process with little sensitivity to the vacancy concentration. For endothermic materials, such as tetracene, the impossibility of such vibronic resonances renders fission fully incoherent; a process that is shown to slow down with vacancy concentration.

## Introduction

Singlet fission (SF), the spontaneous decay of a photoexcited singlet exciton into two triplet excitons in organic molecular materials, poses a fundamental conundrum as well as a promising avenue for the optimization of photovoltaics and spintronics[1-4]. Great strides have been made in improving our mechanistic understanding of the SF process. A spin-correlated pair of triplets has been recognized as an important SF intermediate[4-6], by rendering the process spin allowed, as well as charge transfer states that directly mix with the singlet and triplet pair states[7-12], causing the process to effectively entail the transfer of a single electron. Among other factors that have been deemed important are an approximate energetic resonance between the singlet and triplet pair states as well as discrete vibrational modes that match energetic offsets between these states[9,13-20].

These efforts notwithstanding, a systematic understanding of SF remains lacking, in particular how the aforementioned factors synergistically steer SF under both exothermic and endothermic conditions. Crystalline pentacene (PEN, $C_{22}H_{14}$) has been widely considered as a successful prototypical SF material, where the process is markedly exothermic, efficient, and prompt[21,22]. However, a variety of endothermic materials have been shown to reach a similar SF effectiveness while not suffering from a loss of energy in the singlet-to-triplet conversion event. The prototypical endothermic SF material is crystalline tetracene (TET, $C_{18}H_{12}$), which among SF materials exhibits triplet excitons that are best suited for applications in conjunction with silicon-based photovoltaics[23-25]. However, the precise manner in which different energetic alignments (exothermic versus endothermic) affect SF kinetics and yields has yet to be determined. Another aspect of SF that continues to be a topic of debate is the significance of coherent mixing between the singlet and triplet pair states, which was first prompted by experimental work on pentacene in 2011[22], and which has been discussed for various materials ever since[10,26-34]. The presence of coherent mixing is challenging to probe experimentally and is commonly inferred when SF is unusually fast and/or shows a lack of temperature dependence[11,26,28-30,35]. However, such behaviour can potentially reflect other processes[17,36], and complementary approaches to substantiate the importance of coherence in SF are highly desirable.

Here, we show a remarkable difference in the role of coherence between exothermic and endothermic SF in mixed thin films of PEN and TET co-deposited with a high-bandgap organic compound. Such non-interacting spacer molecules effectively form vacancies, the concentration of which allows us to accurately control the number and (average) distance of neighbouring acenes (PEN or TET)[37-39]. We utilize this control as a complementary means to interrogate the coherent behaviour of PEN- and TET-based blends using time-resolved spectroscopy, an approach that is substantiated by accompanying computational modelling that reproduces the key experimental observations and quantifies the degree of coherent mixing in microscopic detail. These prototypical SF materials have been selected to compare the SF process under exothermic and endothermic conditions. For exothermic PEN-based blends, the SF rate is found to be robust against variations in the concentration of spacer molecules[38], being instead correlated with the (weakly varying) degree of coherent mixing. In marked contrast, the SF rate of endothermic TET-based blends is found to monotonically decrease upon adding spacer molecules, being correlated with the average number of available neighbouring TET molecules through which it engages in incoherent SF. We thus find a presence and absence of coherent SF in PEN and TET, respectively, which is rationalized by the possibility and impossibility of energetic resonances between the relevant singlet and triplet pair states generated by vibrational modes under exothermic and endothermic conditions.

## Results

**Structural and optical characterization of the blends**

Similar to PEN, neat TET deposited on weakly-interacting substrates such as glass grows in a herringbone arrangement with the long molecular axis almost perpendicular to the substrate plane[40], such that each molecule has 4 neighbours in the plane parallel to the substrate where the intermolecular interactions are strongest. The non-interacting spacer molecules used as vacancies in our TET blends were picene (PIC, $C_{22}H_{14}$) and [6]-phenacene (6PH, $C_{26}H_{16}$); see Fig. 1a. We performed a combined structural and optical characterization of the blends in order to confirm statistical intermixing of TET and spacer molecules and to exclude co-crystal formation or phase separation. The in-plane unit cell parameters $a$ and $b$, obtained by X-ray diffraction, are shown in Fig. 1b,c and the Supporting Information. With a decreasing concentration of TET molecules both parameters increase continuously indicating statistical mixing with a uniform random occupation of lattice sites by either TET or spacer molecules. For TET fractions exceeding 50% we observe limited intermixing[41,42] leading to a phase separation between neat TET domains and mixed domains. For this reason, we focussed our experimental analysis on blends with TET fractions below 50%. Within this range, the variations in unit cell parameters are very small, suggesting that the electronic couplings between TET molecules are comparable across the relevant range of TET fractions.

For comparison, the in-plane unit cell parameters of the complementary system PEN:PIC[38,41] are also shown in Fig. 1b,c exhibiting a behaviour similar to the TET-based blends. For all blends, the variation of the (larger) out-of-plane lattice spacing $d$ with the TET or PEN fraction is shown in Fig. 1d.

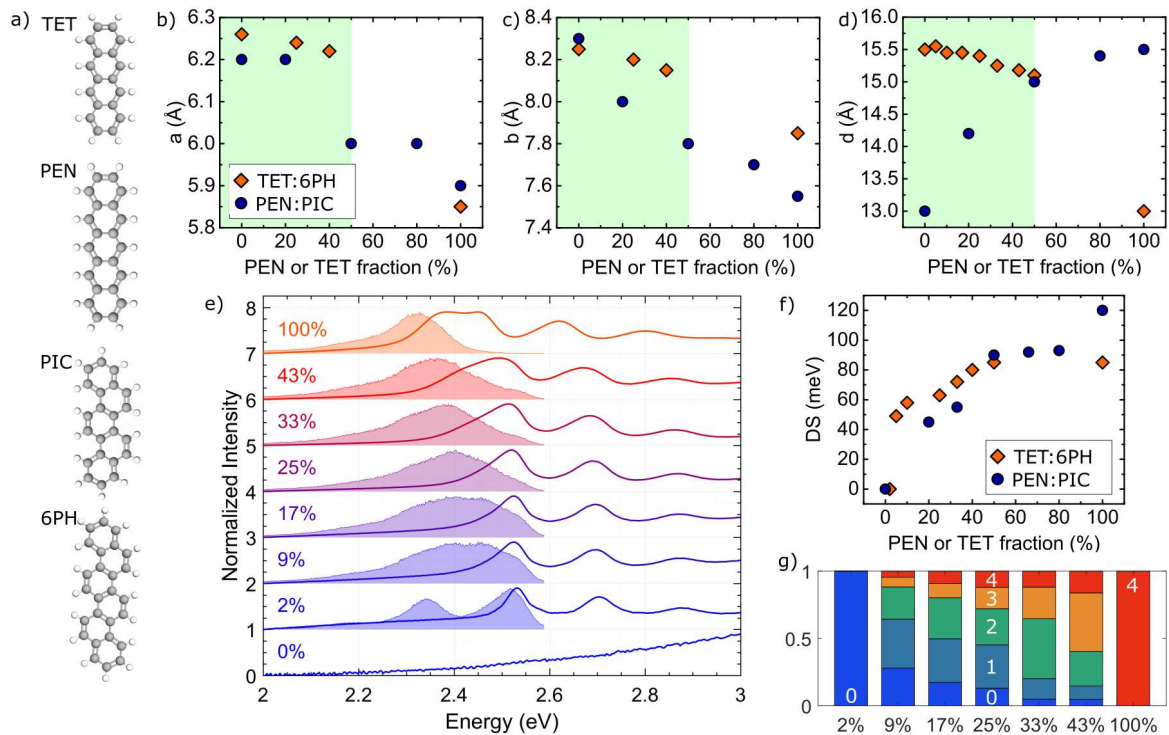

*Figure 1: Structural and optical properties of TET:6PH blends and comparison with data for PEN:PIC blends taken from Ref. 38. (a) Chemical structures of TET, PEN, PIC and 6PH. (b) – (d) In-plane unit cell parameters a and b and out-of-plane lattice spacing d for blends with varying concentrations of spacer molecules. The areas shaded in light green denote the range of PEN[38] or TET fractions relevant for this study. The legend in (b) applies to (b)-(d). (e) Absorption spectra (solid lines) and photoluminescence (PL) spectra (shaded areas) of TET:6PH blends (offset for clarity). (f) Davydov splitting (DS) obtained by fitting the absorption spectra in the relevant photon energy range (2.35 eV – 2.6 eV). For comparison, the DS of PEN:PIC blends[38] is also shown. (g) Decomposition of the PL spectra with differing TET fractions into five contributions depending on the number of TET neighbours for a given TET molecule (see numbers in the bars). The monomer contribution corresponds to 0 TET neighbours while the bulk contribution corresponds to 4 TET neighbours. For details, see text.*

More information on the electronic properties of the blends is provided by linear absorption spectroscopy. Shown in Fig. 1e are absorption spectra taken for the TET:6PH blends in the 2.0-3.0 eV range, which can be unambiguously assigned to the electronic transitions of TET molecules since 6PH has a bandgap of 3.1 eV. With decreasing TET fraction, the shape of the absorption band evolves from the neat TET thin film spectrum to the monomer spectrum, with distinct peaks that are blueshifted by about 100 meV at a TET fraction of 2%, indicating that the majority of TET molecules have become completely isolated. Besides the blueshift due to changes in the polarizability of the molecular environment, the main spectral change with decreasing TET fraction is manifested in the two Davydov components, which are the two lowest-lying absorption peaks located at 2.35 eV and 2.45 eV for neat TET. The energetic separation between these components, the Davydov splitting (DS), results from electronic interactions between the two translationally inequivalent molecules in the unit cell. In going from 50% to 5% TET fraction the DS shown in Fig. 1f is seen to decrease monotonically from about 85 meV to 50 meV, indicating a decrease in the admixture of charge transfer states in the lowest-lying electronic transition[12]. However, this decrease is much less pronounced than in analogous PEN-based blends (Fig. 1f and Ref. 38) which is consistent with the comparatively weaker admixture of charge transfer states in TET[7,12].

Also shown in Fig. 1e are the steady-state photoluminescence (PL) spectra taken for the TET:6PH blends, showing a continuous change of the spectral shape from neat TET to that of isolated monomers. Due to the statistical distribution of spacer molecules within the film, TET molecules in the blend have a varying number of other TET molecules as in-plane neighbours ranging from 4 (the bulk limit) down to 0 (the monomer limit), with the relative contribution of these configurations changing with TET fraction. These variations are borne out in Fig. 1g where the PL spectrum is deconvolved into spectral contributions from such configurations (see Supporting Information for details).

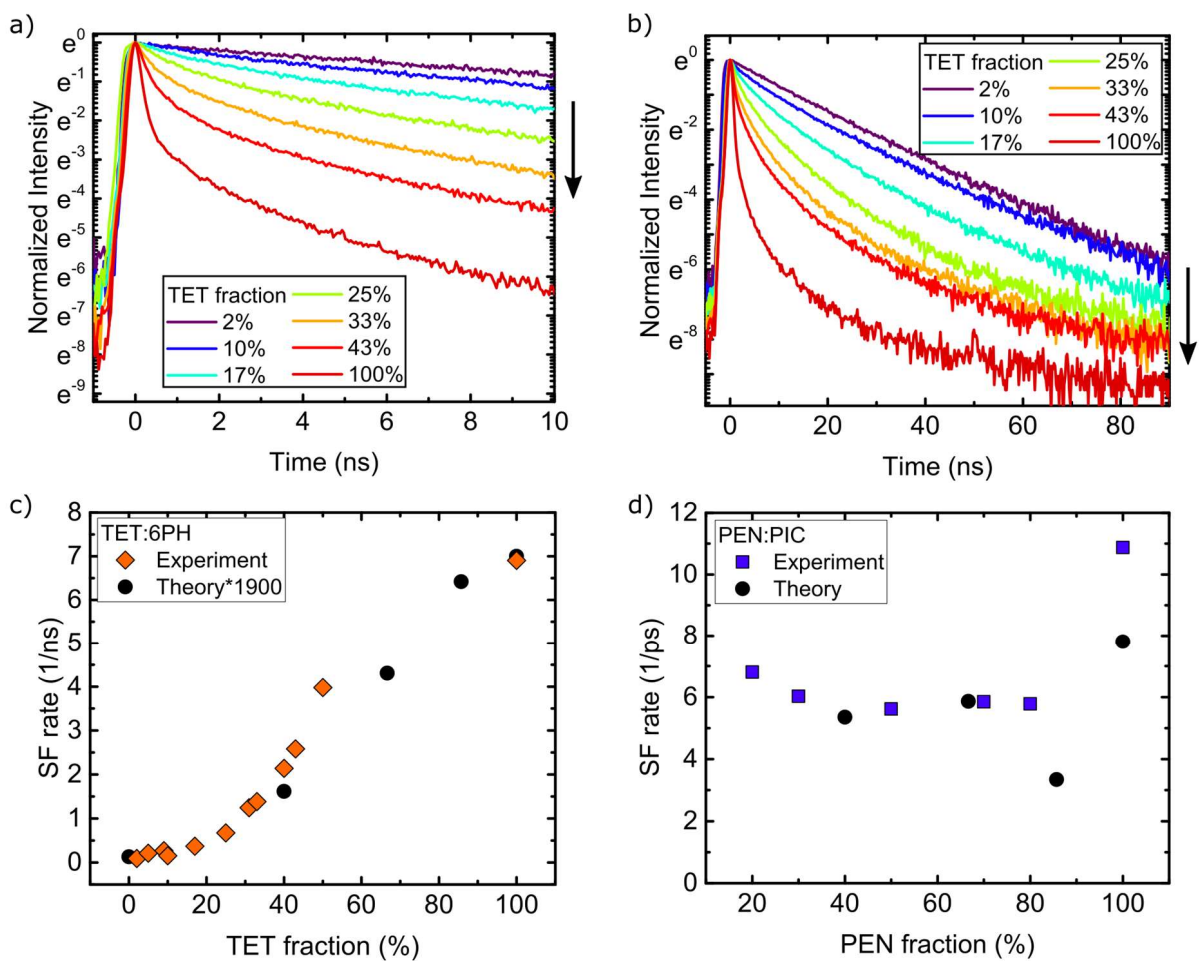

*Figure 2: Time-resolved photoluminescence (TRPL) measurements and comparison of measured and calculated SF rates. (a) TRPL intensity of TET:6PH blends with different TET fractions in a 10 ns window. (b) Equivalent to (a) but in a 90 ns window. The arrows in (a) and (b) denote the direction of increasing TET fraction. (c), (d) Comparison of experimentally (symbols in color) and theoretically (black circles) determined SF rate constants for TET- and PEN-based[38] blends with varying concentrations of spacer molecules. For TET:6PH blends (c) the SF rate has been calculated by subtracting the monomeric decay rate (determined by fitting the 2% TET blend) from the decay rate of the prompt fluorescence, as explained in the text. The experimental data of PEN:PIC blends (d) are adapted from Ref. 38.*

**Time-resolved spectroscopy and determination of singlet fission rates**

In order to determine the SF dynamics in the TET:6PH blends we performed time-resolved photoluminescence (TRPL) spectroscopy. The TRPL intensity shown in Fig. 2a,b is indicative of the time-dependent photoexcited singlet state population, and features a prompt, sub-nanosecond decay followed by a low-amplitude, longer-lived tail. These fast and slow components have in previous reports on neat TET thin films been assigned to SF and delayed fluorescence from triplet-triplet annihilation, respectively[23,43]. At high TET loading, the latter attribution is confirmed by its magnetic field dependence. For decreased TET concentrations, however, the slow component does not change with applied magnetic field (see Supporting Information), indicating that it is not due to enhanced TTA[44], and we instead assign it to singlet state emission from fully-isolated TET monomers within the blends. Importantly, the fast component is seen to dramatically slow down with decreasing TET fraction. In the following, we focus on this component, as it reflects the lower limit of the concentration-dependent SF rate.

Based on the shape of the PL spectrum and the monoexponential decay of the PL intensity we assume that the dynamics of the blend with the lowest TET fraction (2%) is dominated by isolated TET molecules, and that we can extract the concentration-independent monomeric decay rate constant ($k_{monomer}$) from a fit of this data[45]. For the blends with higher TET fraction ($f$), the total rate constant of the prompt decay is given by $k_{SF}(f) + k_{monomer}$ where the fraction-dependent contribution $k_{SF}(f)$ can be assigned to SF.

The fraction-dependent SF rate constants, $k_{SF}(f)$, obtained by fitting the prompt decay of the TRPL measurements, are shown in Fig. 2c. Starting from a TET fraction of 50%, we observe a steep drop of the SF rate with decreasing TET fraction from 7 ns$^{-1}$ for neat TET films to 0.3 ns$^{-1}$ for a TET fraction of 10%. Blends of TET with PIC exhibit a similar behaviour (see Supporting Information), from which we conclude that the reduced SF rate is a result of vacancies introduced by non-interacting spacer molecules, and not due to specific TET-6PH interactions.

Discussion

The SF rate of PEN:PIC blends was previously observed to be insensitive to the concentration of spacer molecules (Fig. 2d and Ref. 38), which was rationalized by fast hopping of singlet excitons to low-energy "hot spots" where SF proceeds effectively[38,46]. In this regard, the strong concentration dependence of the SF rate found for TET-based blends is surprising, as the comparatively lower SF rate constants for TET in combination with high hopping mobilities[47] should enable even more excitons to reach such hot spots. It is unlikely that the SF rate reduction with introducing spacer molecules stems from a reduced interaction strength between TET molecules, since the changes in unit cell parameters and DS are smaller than found for the insensitive PEN blends. We thus find an unexpected contrast in vacancy-dependent SF between the exothermic and endothermic blends, which potentially holds information on the mechanistic principles steering this process.

To obtain a microscopic insight into the SF dynamics for the TET-based and PEN-based blends, we performed calculations employing the theoretical model previously used in detailed studies on neat PEN crystals[9,17], here extended to include vacancies. In short, the acenes are represented using a limited selection of electronic states, including the (singlet) ground and first singlet excited states, ionic (electron) and cationic (hole) states, and the first triplet state, each of which is linearly coupled to a single intramolecular vibration ($\omega_0$). The dynamics under the associated Holstein Hamiltonian is calculated by Markovian Redfield theory in the secular approximation, assuming a Debye spectral density for the vibrational modes other than $\omega_0$ (for details, see Supporting Information and Refs. 9 and 17). The parametrizations of the TET and PEN thin films are based on previous theoretical studies[7,17]. The vacancies are considered as empty sites that are distributed over the crystal lattice such that the average number of neighbouring molecules is taken to be consistent with the expected value for each of the vacancy concentrations. This approach obviously does not involve phase separations between mixed and neat TET domains, and we can utilize the entire range of concentrations to study the SF process. The unit cell parameters and intermolecular couplings were kept constant, allowing us to focus entirely on the effect of vacancies on the SF rates, although future efforts to further elucidate the potential effect of unit cell parameter variations as discussed in Refs. 12 and 38 would be interesting. Lastly, whereas the calculations quantitatively reproduce the rate constants for PEN, we have not addressed the notoriously difficult issue of reaching quantitative agreement for TET (which could be realized by adjusting the Debye spectral density), instead focusing on the qualitative trends.

Calculations of the SF dynamics for TET-based and PEN-based blends, shown in Fig. 3a, show behaviour consistent with the experimental observations: a strong vacancy-dependence of the SF rate for TET, and a weak dependence for PEN. The corresponding SF rates, obtained by fittings to an exponential, are shown alongside the measured analogues in Fig. 2c,d and are seen to qualitatively capture the vacancy-dependent trends for both TET and PEN. As such, we can employ our calculations to unravel the microscopic mechanisms underlying these trends, including those that are inaccessible by experiments alone.

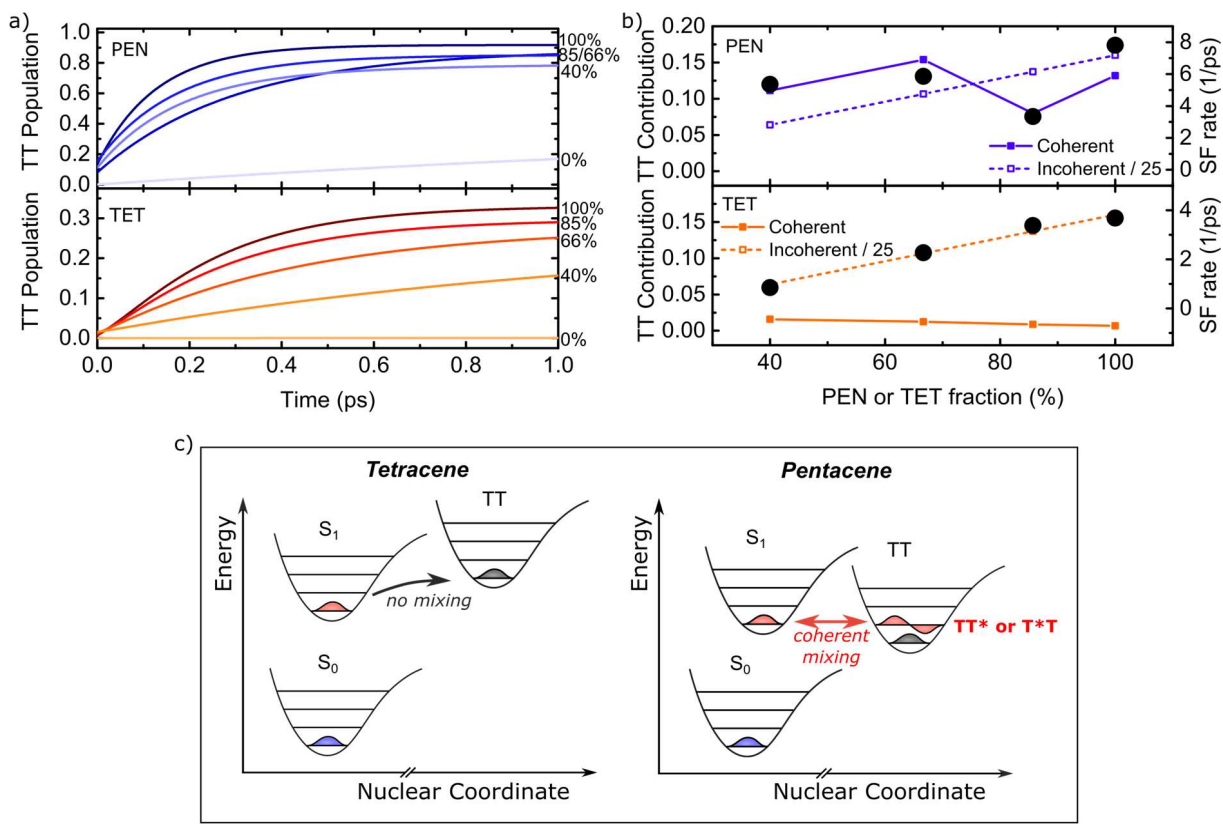

*Figure 3: Triplet pair state population and SF mechanisms. (a) Calculated dynamics of the triplet pair (TT) state population with time for PEN and TET. (b) Contributions of the coherent (filled symbols, solid lines) and incoherent (open symbols, dashed lines) pathways for different mixing ratios for PEN and TET. The corresponding calculated SF rates are shown as black circles. All values at 0% PEN or TET fraction vanish for trivial reasons and are not shown here. For details, see text. (c) Schematic of the two pathways of SF in PEN and TET. Due to the endothermicity of SF in TET, the triplet pairs cannot coherently mix into the photoexcited singlet excited state, in contrast to PEN where such mixing is mediated by a vibrational quantum (denoted by the \*).*

The most pronounced difference between TET- and PEN-based blends borne out in our calculations is the degree of coherent mixing between the triplet pairs and singlet excitations. For the photoexcited singlet state $S_1$ we observe a ~15% triplet pair admixture for neat PEN, against a <1% admixture for neat TET, as shown in Fig. 3b. Moreover, these admixtures are found to be largely independent of the vacancy concentration, being instead dependent on the energy alignment of $S_1$ with the adiabatic triplet pair state TT for which changes are only small. Interestingly enough, for PEN-based blends the small fluctuations apparent in this admixture (related to finite size effects in our calculations) are found to be faithfully followed by the calculated SF rate, which is also shown in Fig. 3b. The observed correlation between the SF rate and the degree of coherent mixing leads us to conclude that coherence between $S_1$ and TT forms the driving force for SF in PEN; a mechanism that is reasonably robust against the introduction of vacancies. In marked contrast, for TET coherent mixing is at least an order of magnitude smaller as a result of which SF is primarily driven by an incoherent mechanism. Consequently, the SF rate is expected to scale as the number of available neighbours for each TET molecule, which is confirmed by the almost perfect correlation between these two quantities in Fig. 3b.

Although the idea of a functional relevance of the coherence between $S_1$ and TT in PEN traces back to 2011[22], the origin of their coherent mixing has been the topic of a longstanding debate. The original hypothesis of electronic couplings between $S_1$ and TT being sufficiently strong to facilitate such mixing[22] has been contradicted by theoretical calculations[48-51]. Indeed, this hypothesis would be at odds with our findings showing an order of magnitude smaller mixing for TET despite similar coupling strengths. More recently, various studies have suggested the presence of a vibrational resonance between $S_1$ and TT to facilitate the sub-100 fs SF time constant in PEN[14,15,17]. In particular, recent theoretical work[17] has shown an intramolecular vibrational mode with an energy of $\omega_0$=1150 cm$^{-1}$ to be responsible for this vibronic enhancement. Although not emphasized at the time, from the data in that work it can be seen that $\omega_0$ not only modulates the SF rate constant, but also the degree of admixture of triplet pair states into $S_1$.

This brings us to the important observation that the functionally-important vibronic resonance generated by $\omega_0$ appears to be unique to exothermic SF, where the TT state dressed by a vibrational quantum (denoted TT*) becomes resonant with the photoexcited state $S_1$, as illustrated in Fig. 3c. For endothermic SF, TT lies already above $S_1$, and instead there only could be a vibronic resonance between the "hot" singlet state $S_1$* and TT. Importantly, this principle leaves the relaxed photoexcitation $S_1$ without a resonant triplet pair state, which explains why coherent mixing is much smaller for TET than for PEN and why, consequently, SF proceeds through an incoherent mechanism instead.

## Conclusion

In summary, our combined experimental and theoretical study shows that exothermic SF is susceptible to a coherent driving mechanism induced by vibrational resonances between the photoexcited singlet state and product triplet pair states. For endothermic materials, no such vibrational resonances are possible, as a result of which SF proceeds incoherently. These mechanisms are at the core of the surprising contrast between SF in PEN-based and TET-based thin films, which are invariant to and variable against introducing spacer molecules, respectively. As such, the results obtained for these prototypical materials provide a comprehensive scenario of how SF results from the interplay of electronic energy alignment, vibronic coupling, and quantum coherence, and offer design rules for devising SF materials. For example, the desirable property of endothermicity, which enhances the energy efficiency of the process, comes at the cost of a heightened susceptibility to material defects. Exothermic SF, on the other hand, has a high degree of robustness, at the cost of excess energy deposited in vibrational modes. Hence, which scenario is more desirable depends on the energy and robustness requirements that are called for.

Progress in understanding the role of coherence in SF has been hampered by the complexity of the SF process as well as the difficulty of detecting quantum coherence. In the present study, we complement commonly-used time-resolved optical measurements of SF with the controlled introduction of vacancies in both exothermic and endothermic settings, in order to fundamentally expand the number of tuneable parameters with which to interrogate the SF process. The contrasting behaviour of exothermic PEN and endothermic TET with varying vacancy concentration is a telltale of the complexity of SF, as its explanation involves the simultaneous involvement of energy alignment, vibronic coupling, and quantum coherence. The difficulty of detecting coherence has been overcome by accompanying calculations that simultaneously reproduce the key experimental observations and allow the degree of coherent mixing to be quantified. Hence, by combining comprehensive experimental control with theoretical modelling, we have followed a holistic approach that we anticipate to find repeated success in disentangling a broad range of complex photophysical processes.

## Methods

**Sample preparation**

The mixed thin films of TET (Sigma Aldrich, 99.99%) and 6PH (Lambson Japan Co. Ltd. 99%) or PIC (Tokyo Chemical Industry Co. Ltd. 99.9%) were grown by organic molecular beam deposition on silicon with a native oxide layer and on borofloat glass substrates at a base pressure of $1\times10^{-9}$ mbar. The total growth rate was 0.6 nm/min, with the rates of the two materials monitored separately by two quartz crystal microbalances, calibrated using X-ray reflectivity. Unless noted otherwise, the final film thickness was 80 nm.

**Structural and optical characterization**

X-ray reflectivity was measured on a diffractometer (3303TT, GE) using Cu Kα-radiation ($\lambda = 1.5406$ Å) and a 1D detector (Meteor 1D, XRD Eigenmann). Grazing incidence X-ray diffraction and the reciprocal space maps were measured at beamline SixS at Soleil, Gif-sur-Yvette Cedex, France using a wavelength of $\lambda = 0.9538$ Å. The sample was kept in vacuum during the measurements to avoid beam damage.

UV-vis transmission spectra were measured using a Perkin Elmer Lambda 950 spectrophotometer. Photoluminescence excitation spectroscopy was performed on a Horiba Fluorolog-3 DF spectrofluorimeter using a 450 W Xenon lamp for excitation and a Hamamatsu R2658P PMT to monitor the emission.

**Time-resolved photoluminescence**

Time-resolved photoluminescence measurements were taken with a Hamamatsu C4334 Streakscope having a time resolution of 20 ps and a spectral resolution of 2.5 nm. The 800 nm output of an 80 MHz Coherent Vitesse Ti:Sapphire oscillator was frequency doubled to generate the 400 nm excitation pulse. A Pockels cell controlled by a ConOptics pulse picking system was used to adjust the repetition rate of the oscillator to 100 kHz. A 450 nm long wave pass filter was placed before the streak camera to minimize the contribution of laser scatter to the signal. All measurements were performed in a vacuum cryostat ($10^{-5}$ torr) fitted with optical windows, and pulse fluences remained below 1.2 μJ/cm$^2$.

**Modeling**

Crystalline tetracene and pentacene were parametrized based on previous theoretical studies[7,9,17]. A full microscopic basis was used including for each acene molecule a (diabatic) singlet ground and a singlet excited state as well as a cationic, an anionic and a first triplet state. Moreover, a single quantum harmonic oscillator was included for each molecule to account for linear coupling of the electronic transitions to a single intramolecular vibration by means of the Holstein Hamiltonian. The quantum dynamics was obtained by applying non-Markovian Redfield theory in the secular approximation using a Debye spectral density to account for the Holstein-coupled vibrational modes other than the quantum harmonic oscillator. The non-interacting spacer molecules were accounted for by taking out acene molecules and their associated electronic states from the microscopic basis set.

## Acknowledgements

We thank Frank Schreiber for access to equipment, the Soleil Synchrotron for beamtime allocation and Alina Vlad, Rupak Banerjee, Lisa Egenberger and the staff of beamline SixS for support during the synchrotron experiments. This work was supported by the National Science Foundation grant CHE-1800187 (C.J.B.) and NSF-CHE -1464802 (D.R.R.), and the German Research Foundation grant HI 1927/1-1 (J.H.).